Costantino Sigismondi
*Università di Roma
"La Sapienza"*
sigismondi@icra.it

# Incontri celesti, vita del padre Clavio in cinque atti

**Abstract:** 2012 will be the fourth centennial year of the Jesuit Christopher Clavius (1535-1612), known as the Euclid of XVI century and the collaborator of the Pope Gregory XIII for the calendar reformation. In the occasion of the year of astronomy I wrote a short theatre pièce *"Celestial encounters"* dedicated to the life of Ft. Clavius. He observed two total eclipses from centreline in 1560 in Coimbra and in 1567 in Rome, a fact which is remarkable even for contemporary astronomers. The story is developed around those trips: scientific and religious motivations are put in evidence with historical and fantasy, but realistic, facts. An interregional project between Switzerland and Italy, dedicated to the development of high resolution CMOS camera for astronomy and medical sciences has been entitled to Clavius and will produce high resolution measurements of solar diameter.

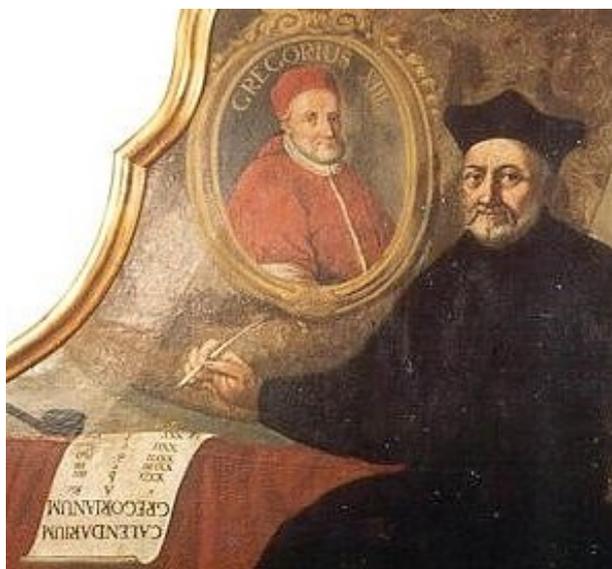

Padre Cristoforo Clavio a fianco del ritratto di papa Gregorio XIII, mentre scrive il trattato sul Calendario Gregoriano (1603).

**Clavio ed il mondo culturale alla fine del '500**

Gesuita bavarese, ha insegnato matematica e astronomia al Collegio Romano per mezzo secolo, illustrando Roma con la sua presenza.

In un anno astronomico nato a celebrare i 400 anni dalle prime osservazioni al telescopio di Galileo, è importante avere presente anche gli altri grandi personaggi che hanno contribuito allo sviluppo di questa che è la regina delle scienze, non fosse altro che per la sua maggiore antichità.

In particolare a cavallo tra il XVI ed il XVII secolo non una, ma molte rivoluzioni erano in corso di svolgimento. Il Padre Clavio (1538-1612) presiedette la commissione per il Calendario, istituita da Gregorio XIII Boncompagni, che portò alla riforma del 1582.

Questa riforma del calendario, insieme ai fini liturgici soddisfaceva anche a quelli scientifici, e politici in quanto contribuì a consolidare la centralità del romano pontefice di fronte ad un cristianesimo che aveva appena sperimentato lo scisma di Lutero.

Della riforma gregoriana Clavio fu il principale "apologeta" e la biblioteca conserva una copia del 1603 sul calendario gregoriano, donata da Clavio al Cardinale Cesare Baronio (1538-1607), il più illustre tra i bibliotecari della Vallicelliana.

Il Baronio, il cui stemma è riprodotto a fianco sulla coperta del libro del 1603 di Clavio *Romani Calendarii a Gregorio XIII Restituti Explicatio* [1], fu un gigante nella storia della chiesa, con la pubblicazione degli Annales ecclesiastici e la revisione del martirologio romano, arginò le affermazioni della storiografia luterana con prove e documenti. Benedetto XIV Lambertini lo proclamò venerabile nel 1745 ed ora Benedetto XVI ha riaperto la sua causa di beatificazione per la gioia degli abitanti di Sora che lo chiamano il Santo senza candele... Il suo motto *Oboedientia et Pax* fu scelto anche da papa Giovanni XXIII. Baronio a sua volta era ispirato da un altro gigante, san Filippo Neri (1515-1595), fondatore degli Oratoriani e pure direttore spirituale di S. Camillo de Lellis (1550-1614), fondatore dei chierici regolari ministri degli infermi che rivoluzionarono il concetto di ospedale ed il servizio ai malati. Benedetto XIV, che aveva incoraggiato l'Accademia delle Scienze di Bologna e ritrovato l'Obelisco del Sole al Campo Marzio (1748) canonizzò proprio S. Camillo. Il Padre Clavio, da par suo, era ritenuto l'Euclide del XVI secolo, gigante anche lui nel campo delle matematiche.

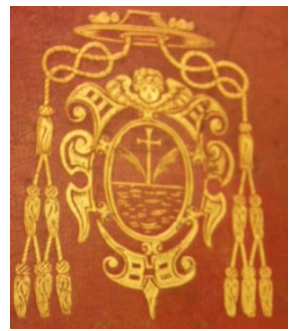

A fianco: dall'anti-porta del testo sugli Elementi di Euclide di Clavio la figura di Euclide[2].

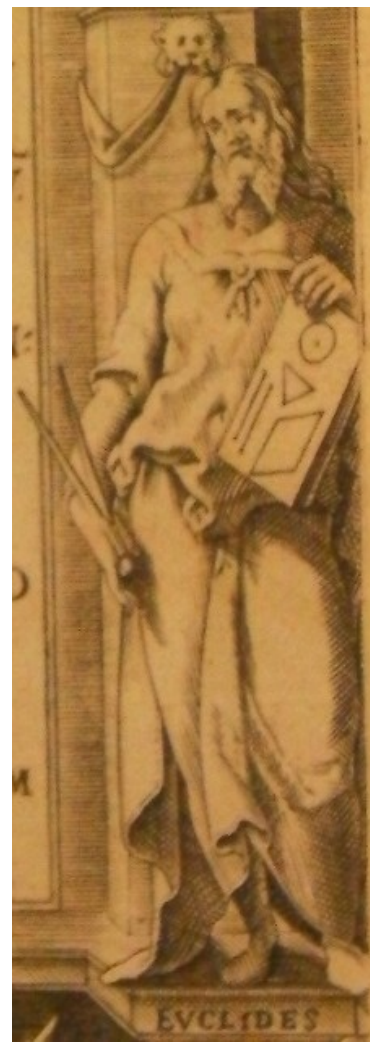

Clavio era giunto a Roma dopo gli studi in Portogallo a Coimbra, in una nazione che dominava il mondo grazie all'arte della
navigazione e ad un carisma particolare nel commercio e nelle relazioni internazionali. Dopo il trattato di Tordesillas (1494), infatti, il Portogallo si era assicurato il dominio delle rotte verso l'oriente che circumnavigavano l'Africa dopo un'ampia svolta ad occidente per seguire i venti favorevoli, come Vasco de Gama fece nel suo viaggio in India (1497-1499). Le caravelle avevano un carico utile cento volte superiore a quello delle carovane che tra grandi difficoltà dovevano attraversare regioni infestate da predoni, per cui la via aperta da Vasco de Gama divenne una grande fonte di ricchezza e Lisbona la prima porta dell'oriente. Tanti gesuiti, tra cui il padre Matteo Ricci (1552-1610, Li Madou per i Cinesi), allievo di Clavio, portarono la scienza europea nell'estremo oriente.

**Clavio e le eclissi**
Clavio osservò a Coimbra l'eclissi totale di Sole del 21 agosto 1560. L'eclissi durò lo spazio di un Miserere (il salmo 50) cioè 3 minuti e 20 secondi, quanto la recita di questo salmo richiede. Clavio descrive nel suo commentario alla Sfera del Sacrobosco [3] di cui la Biblioteca Vallicelliana possiede l'edizione del 1585, l'apparizione delle stelle durante l'eclissi, così come gli uccelli a terra e che non riusciva a vedere i suoi piedi per quanto era buio.
A Roma, nel 1567 ne vide un'altra che al suo massimo lasciò il Sole circondato da un anello sottile di luce, che successivamente, per ben due volte, rispondendo a Keplero, descrisse come un anello e non un'atmosfera attorno al corpo oscuro della Luna. Questa fu per Clavio una prova che il diametro del Sole poteva essere maggiore di quello della Luna, contrariamente a quanto Tolomeo (II secolo d.C.) aveva riportato nell'Almagesto [4], il più autorevole libro di astronomia fino ad allora.

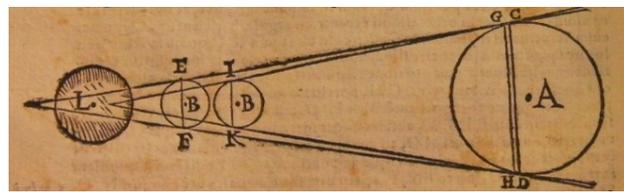

Clavio, *In Sphaeram Ioannis de Sacrobosco* (1585) [3], p. 440 e 442 (figura successiva): in A c'è il Sole, in B e B' la Luna al perigeo e all'apogeo. Si vede che in entrambi i casi il cono d'ombra tocca la Terra (L), in accordo con i dati presentati nell'Almagesto di Tolomeo [Libro V cap. 14], tuttavia Clavio stesso presenta il caso di un'eclissi solare in cui il cono d'ombra non tocca la Terra, quella del 1567 dove vide un anello attorno alla Luna. Durate differenti delle eclissi lunari sono previste con il Sole al perigeo e all'apogeo (immagine seguente).

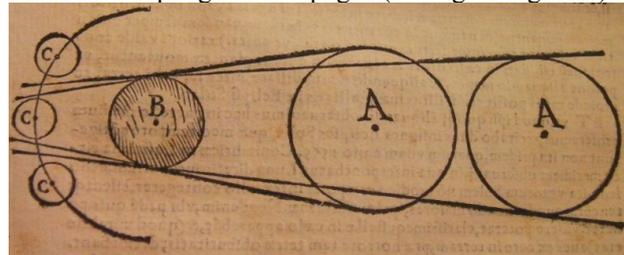

Proprio questa cronaca dell'eclissi del 9 aprile 1567, che l'ultima eclissi totale osservata da Roma fino ad oggi, ha riportato gli scritti di Clavio sul tavolo di molti scienziati contemporanei, che si occupano delle variazioni secolari del diametro del Sole, con le sue conseguenze in climatologia. Se infatti quello che Clavio osservò fosse stata proprio la fotosfera solare avremmo avuto, nel 1567, un Sole circa 6000 Km più grande di quanto non sia oggi. Questa variazione è difficile da concepire nei modelli attuali per il nostro Sole, tuttavia esistono molti studi interessanti scaturiti ancora quattro secoli dopo questa storica osservazione di Clavio [5,6].

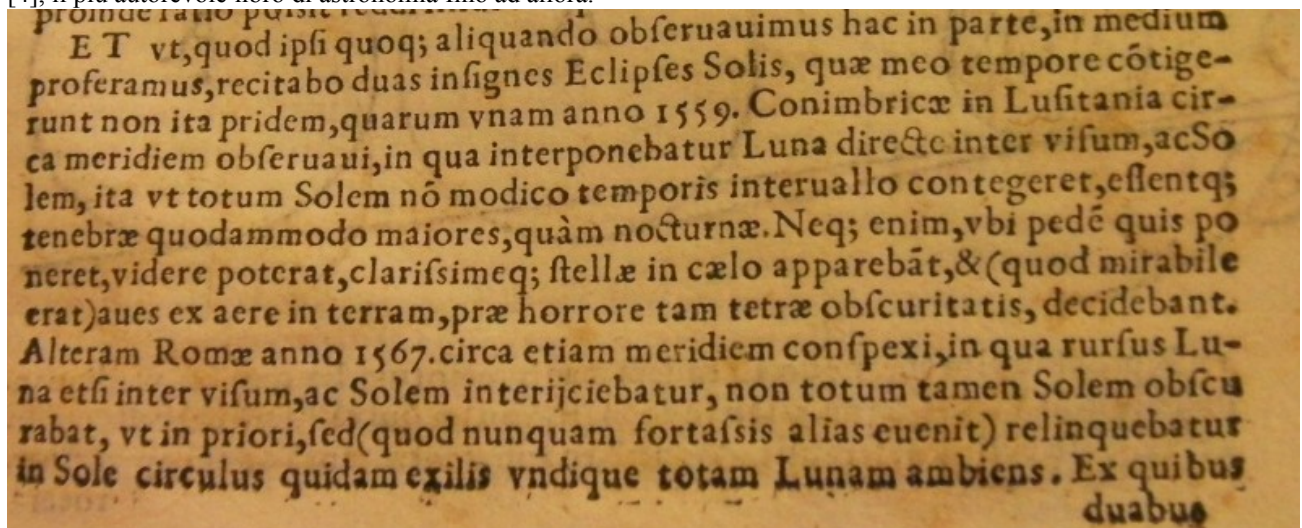

Clavio, *In Sphaeram Ioannis de Sacrobosco* (Roma, 1585), p. 441 [3]. Passo contenente la citazione delle eclissi osservate da Clavio nel 1560 e nel 1567.

Si noti che l'eclissi di Coimbra viene collocata nel 1559, anziché nel 1560. Non deve stupire questa imprecisione dal momento che il Saros, un ciclo di 18 anni 11 giorni e 8 ore circa in capo al quale si ripetono le eclissi fu scoperto solo da Halley. Quindi Clavio non aveva modo di ricontrollare il dato con precisione dai calcoli. E nel 1559 i calcoli mostravano comunque altre eclissi.

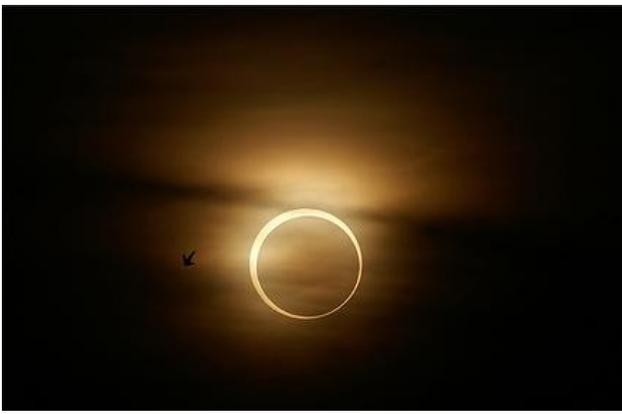

Eclissi anulare nell'Isola di Giava, foto del 26 gennaio 2009. Con il Sole vicino al perielio e la Luna prossima all'apogeo, questa eclissi è stata quella in cui il rapporto tra i diametri Luna:Sole è stato più piccolo e pari a 0.92. I parametri orbitali conosciuti da Tolomeo non consentivano eclissi anulari [3]. Benché a Costantinopoli fu osservata un'eclissi anulare il 22 dicembre 968, riportata in una cronaca del tempo[7], Clavio fu il primo a pubblicare un'osservazione di questo tipo.

Clavio è anche uno dei padri della gnomonica, avendo pubblicato diversi libri sull'argomento degli orologi solari di tutte le forme.

Clavio è stato l'ultimo dei grandi astronomi tolemaici. Negli ultimi anni della sua vita conobbe e stimò Galileo, poté osservare anche i satelliti di Giove, e la scuola che aveva fondato portò frutti copiosi fino in estremo oriente. Sulla Luna gli è stato dedicato uno dei più grossi crateri presso il polo Sud.

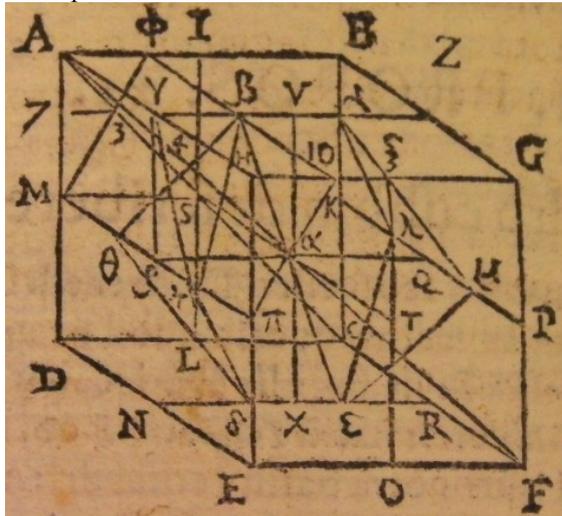

Il dodecaedro (immagine è tratta dall'edizione degli Elementi di Euclide [2] del 1589, tradotta poi anche in cinese) è l'ultimo dei solidi regolari, cioè costruiti con figure piane, a loro volta regolari. Keplero riprese, sei anni dopo l'edizione di questo libro, i solidi regolari per risolvere il *Mysterium Cosmographicum* del numero e delle distanze tra i pianeti visibili nel sistema Copernicano. Seguendo questo metodo deduttivo, del tutto opposto a quello galileiano, scoprirà quella che oggi è nota come la sua terza legge.

**Incontri celesti**
Il titolo della *pièce* è pensata per essere rappresentata con il teatro delle ombre cinesi. Le scenografie sono facili da realizzare sfruttando un lenzuolo ed un proiettore davanti al quale sia disposto un profilo intercambiabile.

Gli attori, idealmente bambini delle scuole elementari, si muovono dietro la quinta, proiettando la loro ombra sul lenzuolo, mentre i lettori proclamano il testo leggendo ciascuno la sua parte.

Gli *incontri celesti* sono quelli eccezionali tra Sole e Luna a cui Clavio assiste nella sua vita di giovane gesuita, ma anche quello con un pastorello in Cova de Irìa nella veglia dell'Assunta, quello con il Papa Gregorio, o quello con l'opera di Gerberto a Magdeburgo che lo orienta già da ragazzo verso lo studio di due delle sette colonne della Sapienza [Libro dei Proverbi **9**, 1] quelle della Geometria e dell'Astronomia.

Il testo è disponibile al sito
www.icra.it/solar/incontri-celesti.pdf

**Il progetto CLAVIUS**
In onore di Clavio è stato battezzato il progetto italo svizzero Clavius, per la misura del diametro solare in diverse bande spettrali. Con l'uso di camere CMOS, sviluppate all'Università di Como, ad alta velocità montate al telescopio solare gregoriano di Locarno (Canton Ticino nella Svizzera Italiana) si riprenderanno transiti solari su cerchi orari in modo da "congelare" il seeing atmosferico, valutandolo in tempo reale e scorporandolo dalla misura osservata del diametro.

In questo modo si potrà tenere sotto controllo le variazioni del diametro solare, che indicano anche l'attività magnetica del Sole che ha delle conseguenze anche sul clima terrestre.

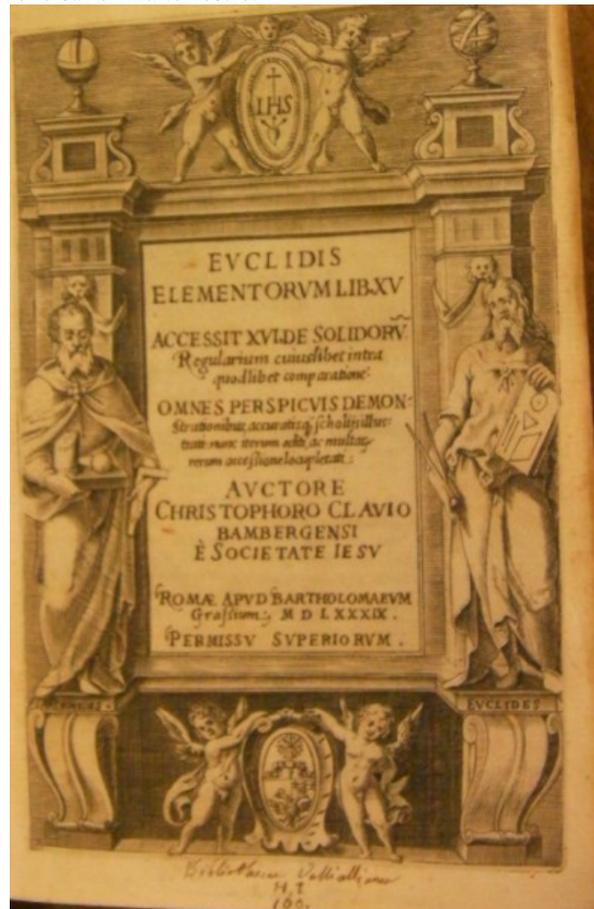

Antiporta del libro su Euclide [2] di Clavio, edizione romana alla Biblioteca Vallicelliana, del 1589.

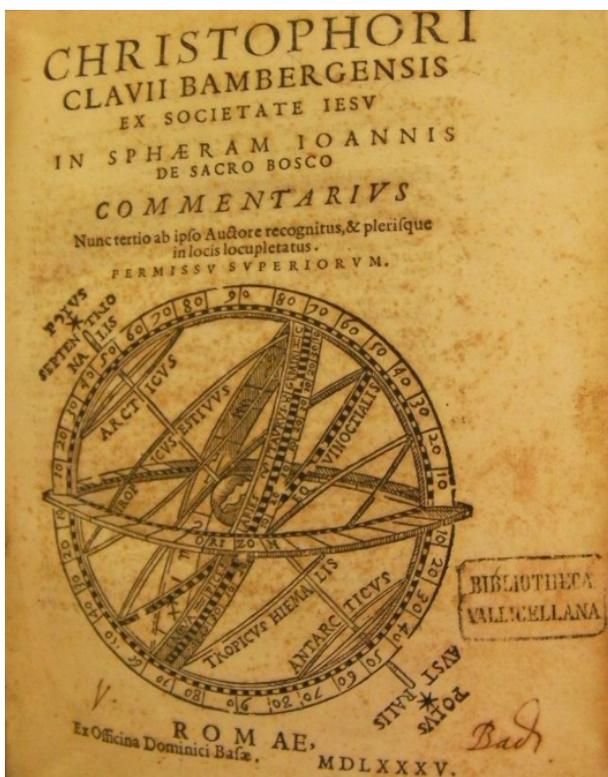

*Commentarius in Sphaeram Ioannis de Sacrobosco*[3], antiporta edizione romana del 1585.

**Crediti:** Le figure in questo testo sono riprodotte su autorizzazione del Ministero per i beni e le attività culturali e sono tratte dai fondi della Biblioteca Vallicelliana in Roma.

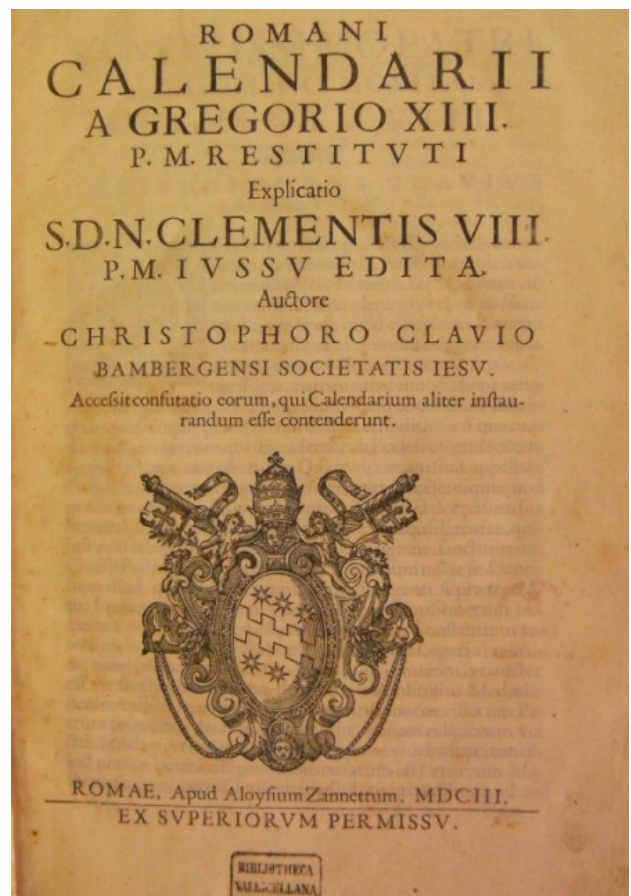

Clavio, *Romani Calendarii a Gregorio XIII Restituti Explicatio* [1], antiporta comprendente lo stemma del papa Clemente VIII Aldobrandini, Roma 1603.